\documentclass[conference]{IEEEtran}
\ifCLASSINFOpdf
\else
  \usepackage[dvips]{graphicx}
\fi
\usepackage{url}

\usepackage{algpseudocode}
\usepackage{algorithm}

\begin{document}
%
\title{LabelRank: A Stabilized Label Propagation Algorithm for Community Detection in Networks}

\author{\IEEEauthorblockN{Jierui Xie}
\IEEEauthorblockA{Department of Computer Science\\Rensselaer Polytechnic Institute\\
110 8th Street\\Troy, New York 12180\\Email: jierui.xie@gmail.com}
\and
\IEEEauthorblockN{Boleslaw K. Szymanski}
\IEEEauthorblockA{Department of Computer Science\\Rensselaer Polytechnic Institute\\
110 8th Street\\Troy, New York 12180\\Email: szymab@rpi.edu}
}


%


\maketitle

\begin{abstract}
An important challenge in big data analysis nowadays is detection of cohesive groups in large-scale networks, including social networks, genetic networks, communication networks and so. In this paper, we propose \textit{LabelRank}, an efficient algorithm detecting communities through label propagation. A set of operators is introduced to control and stabilize the propagation dynamics. These operations resolve the randomness issue in traditional label propagation algorithms (LPA), stabilizing the discovered communities in all runs of the same network. Tests on real-world networks demonstrate that LabelRank significantly improves the quality of detected communities compared to LPA, as well as other popular algorithms.
 
\end{abstract}



%
\IEEEpeerreviewmaketitle

\section{Introduction}
One type of the basic structures of sociology in general and social networks in particular are {\it communities} 
(e.g. see \cite{human-comm}). In sociology, community usually refers to a social unit 
that shares common values and both the identity of the members and their degree of cohesiveness depend on individuals' social and cognitive factors such as beliefs, 
preferences, or needs. The ubiquity of the Internet and social media eliminated spatial limitations on community range, resulting in online communities linking people 
regardless of their physical location. The newly arising {\it computational sociology} relies on computationally intensive methods to analyze and model social 
phenomena \cite{comput-soc}, including communities and their detection. Analysis of social networks 
has been used as a tool for linking micro and macro levels of 
sociological theory. The classical example of the approach is presented in \cite{weak-ties} that elaborated the macro implications of one aspect of small-scale interaction,  
the strength of dyadic ties. Communities in social networks are discovered based on the observed interactions between people. 
With the rapid emergence of large-scale online social networks, e.g., Facebook that connected 
a billion users in 2012, there is a high demand for efficient community detection algorithms that will be able to handle large amount of data on a daily basis. Numerous 
techniques have been developed for community detection. However, most of them require a \textit{global} view of the network. Such algorithms are not scalable enough for 
networks with millions of nodes.

Label propagation based community detection algorithms such as LPA \cite{Raghavan:2007,JieruiXieLPA:2010} and SLPA \cite{JieruiXie-SLPA:2011, JieruiXie-SLPA:2012} (whose source codes are publicly available at \url{https://sites.google.com/site/communitydetectionslpa/}) require only local information. They have been shown to perform well and be highly efficient. However, they come with a great shortcoming. Due to random tie breaking strategy, they produce different partitions in different runs. Such instability is highly undesirable in practice and prohibits its extension to other applications, e.g., tracking the evolution of communities in a dynamic network. 

In this paper, we propose strategies to stabilize the LPA and to extend MCL \cite{MCL:2000} approach that resulted in a new algorithm called \textit{LabelRank} that produces deterministic partitions.  

\section{Related Work}

Despite the ambiguity in the definition of community, numerous techniques have been
developed including Random walks  \cite{ZhouLipowsky:2006, Hu:2008, DBLP:journals/jgaa/PonsL06}, spectral clustering \cite{ShiMalik:2007, WhiteSmyth:2005, Capocci:2005}, modularity maximization \cite{Blondel-2008, PhysRevE.74.036104, PhysRevE.69.026113, WakitaTsurumi:2007, SchuetzCaflisch:2008}, and so on. A recent review can be found in \cite{Santo:2010}. Label propagation and random walk based algorithms are most relevant to our work. 

The LPA \cite{Raghavan:2007} uses the network structure alone to guide its process. It starts from a configuration where each node has a distinct label. At every step, each node changes its label to the one carried by the largest number of its neighbors. Nodes with same label are grouped together after convergence. The speed of LPA is optimized in \cite{JieruiXieLPA:2010}. Leung \cite{Leung:2009} extends LPA by incorporating heuristics like hop attenuation score. COPRA \cite{Gregory:2010} and SLPA \cite{JieruiXie-SLPA:2012} extend LPA to detection of overlapping communities by allowing multiple labels. However, none of these extensions resolves the LPA randomness issue, where different communities may be detected in different runs over the same network. 

Markov Cluster Algorithm (MCL) proposed in \cite{MCL:2000} is based on simulations of flow (random walk). MCL executes repeatedly matrix multiplication followed by inflation operator. LabelRank differs from MCL in at least two aspects. First, LabelRank applies the inflation to the label distributions and not to the matrix $M$. Second, the update of label distributions on each node in LabelRank requires only local information. Thus it can be computed in a decentralized way. Regularized-MCL \cite{kdd-2009-R-MCL} also employs a local update rule of label propagation operator. Despite that, the authors observed that it still suffers from the scalability issue of the original MCL. To remedy, they introduced Multi-level Regularized MCL, making it complex. In contrast, we address the scalability by introducing new operator, conditional update, and the novel stopping criterion, preserving the speed and simplicity of the LPA based algorithms.

\section{LabelRank Algorithm}
LabelRank is based on the idea of simulating the propagation of labels in the network. Here, we use node id's as labels. LabelRank stores, propagates and ranks labels in each node. During LabelRank execution, each node keeps multiple labels received from its neighbors. This eliminates the need of tie breaking in LPA\cite{Raghavan:2007} and COPRA \cite{Gregory:2010} (e.g., multiple labels with the same maximum size or labels with the same probability). Nodes with the same highest probability label form a community. Since there is no randomness in the simulation, the output is deterministic. LabelRank relies on four operators applied to the labels: (i) propagation, (ii) inflation, (iii) cutoff, and (iv) conditional update.

\begin{figure*}	
  \centering
	\includegraphics[width=15cm, height=7.5cm]{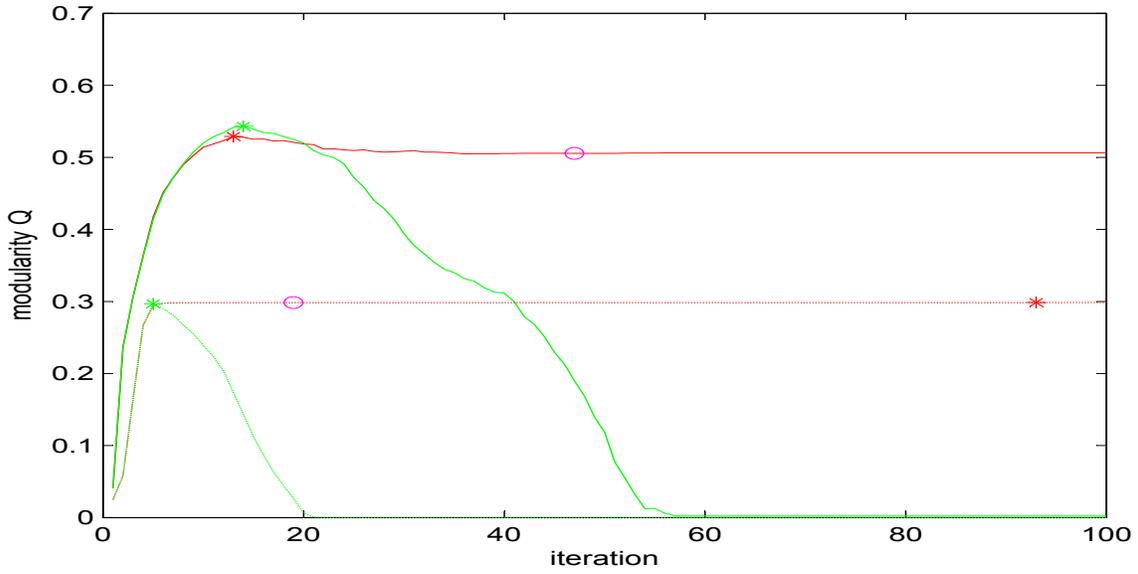}
	\caption{The effect of conditional update operator. The plot shows the modularity $Q$ over iterations on the email network with $n=1,133$ (two curves on the top) and wiki network  with $n=7,066$ (two curves at the bottom). Each $Q$ is computed explicitly for each iteration. Green curve is based on three operators Propagation+Inflation+Cutoff. Red curve is based on four operators Propagation+Inflation+Cutoff+Conditional Update. Asterisk indicates the best performance of $Q(t)$. Purple circle indicates the $Q$ achieved when the stop criterion described in the main text is used.}
	\label{fig:stopcriterion}
\end{figure*}

\textbf{Propagation}: In each node, an entire distribution of labels is maintained and spread to neighbors. We define $n$ $1 \times n$ vectors $P_i$ ($n$ is the number of nodes) which are separate from adjacency matrix $A$ defining the network structure. Each element $P_i(c)$ or $P_{ic}$ holds the current estimation of probability of node $i$ observing label $c \in C$ taken from a finite set of alphabet $C$. For clarity of discussion, we assume here that $C=\{1,2,\dots,n\}$ (same as node id's) and  $|C|=n$. In Section \ref{sec:impdis} we lift this assumption to increase efficiency of execution. In LabelRank, each node broadcasts the distribution to its neighbors at each time step and computes the new distribution $P_i^{'}$ simultaneously using the following equation:
\begin{equation}
\label{eq:bc}
P_i^{'}(c)=\sum_{j\in Nb(i)} P_j(c)/k_i , \forall c\in C,
\end{equation}
where $Nb(i)$ is a set of neighbors of node $i$ and $k_i=|Nb(i)|$ is the number of neighbors. Note that, $P_i^{'}$ is normalized to make a probability distribution. 

In matrix form this operator can be expressed as: 
\begin{equation}
\label{eq:AP}
A \times P,
\end{equation}
where $A$ is the $n \times n$ adjacency matrix and $P$ is the $n \times n$ \textit{label distribution matrix}. To initialize $P$, each node is assigned equal probability to see each neighbor: 
\begin{equation}
\label{eq:initP}
P_{ij}=1/k_i, \forall j \mbox{ s.t. } A_{ij}=1.   
\end{equation}

Since the metric space $A$ is usually compact, $P$ defined iteratively by Eq. \ref{eq:AP} converges to the same stationary distribution for most networks by the Banach fixed point theorem \cite{Banachfixedpoint:1922}. Hence, a method is needed for trapping the process in some local optimum in the quality space (e.g., modularity $Q$ \cite{newman-2004-69}) without propagating too far.  

\textbf{Inflation}: As in MCL \cite{MCL:2000,kdd-2009-R-MCL}, we use the inflation operator $\Gamma _{in}$ on $P$ to contract the propagation, where $in$ is the parameter taking on real values. Unlike MCL, we apply it to the label distribution matrix $P$ (rather than to a stochastic matrix or adjacency matrix) to decouple it from the network structure. After applying $\Gamma _{in} P$ (Eq. \ref{eq:inflation}), each $P_i(c)$ is proportional to $P_i(c)^{in}$, i.e., $P_i(c)$ rises to the $in^{th}$ power.  
\begin{equation}
\label{eq:inflation}
\Gamma _{in} P_i(c)=P_i(c)^{in}/ \sum_{j \in C} P_i(j)^{in}.
\end{equation}
This operator increases probabilities of labels that were assigned high probability during propagation at the cost of labels that in propagation received low probabilities. For example, two labels with close initial probabilities 0.6, and 0.4 after $\Gamma_{in=2}$ operator will changed probabilities to 0.6923 ad 0.3077, respectively. In our tests, this operator helps to form local subgroups. However, it alone does not provide satisfying performance in large networks. Moreover, the memory inefficiency problem implied by Eq. \ref{eq:AP}, i.e., $n^2$ labels stored in the networks, is not yet fully resolved by the inflation operator.

\textbf{Cutoff}: To alleviate the memory problem, we introduce cutoff operator $\Phi_{r}$ on $P$ to remove labels that are below threshold $r\in [0,1]$. As expected, $\Phi_{r}$ constrains the label propagation with help from inflation that decreases probabilities of labels to which propagation assigned low probability. More importantly, $\Phi_{r}$ efficiently reduces the space complexity, from quadratic to linear. For example, with $r=0.1$, the average number of labels in each node is typically less than $3.0$.

\textbf{Explicit Conditional Update}: As shown in Fig.~\ref{fig:stopcriterion} (green curve), the above three operations are still not enough to guarantee good performance. This is because the process detects the highest quality communities far before convergence, and after that, the quality of detected communities decreases. Hence, we propose here a novel solution based on the {\em conditional update operator} $\Theta$. It updates a node only when it is significantly different from its neighbors in terms of labels. This allows us to to preserve detected communities and detect termination based on scarcity of changes to the network. At each iteration, the change is accepted only by nodes that satisfy the following update condition:
\begin{equation}
\label{eq:updatecondition}
\sum_{j \in Nb(i)} isSubset(C^{*}_{i},C^{*}_{j}) \leq qk_i,
\end{equation}
where $C^{*}_{i}$ is the set of \textit{maximum labels} which includes labels with the maximum probability at node $i$ at the \textit{previous} time step. Function $isSubset(s_1,s_2)$ returns 1 if $s_1 \subseteq s_2$, and 0 otherwise. $k_i$ is the degree of node $i$, and $q$ is a real number parameter chosen from the interval $[0,1]$. Intuitively, isSubset can be viewed as a measure of \textit{similarity} between two nodes. As shown in Fig.~\ref{fig:stopcriterion}, $\Theta _{q}$ operator successfully traps the process in the modularity space with high quality, indicated by a long-lived plateau in the modularity curve (red curves). Equation \ref{eq:updatecondition} augments the stability of the label propagation. 

\textbf{Stop criterion}: One could define the \textit{steady} state of a node as small difference in the label distribution between consecutive iterations, and determine the overall network state built upon node states. In fact, the above conditional update allows us to derive a more efficient stop criterion (linear time). We determine whether the network reaches a relatively stable state by tracking the number of nodes that update their label distributions (i.e., implicitly tracking the number of nodes that potentially change their communities), $numChange$, at each iteration and accumulate the number of repetitions $count(numChange)$ in a hash table. The algorithm stops when the $count$ of any $numChange$ first exceeds some predefined frequency (e.g., five in our experiments), or no change for this iteration (i.e., numChange=0).

Although such criterion does not guarantee the best performance, it almost always returns satisfying results. The difference between the found $Q$ (purple circles) and maximum $Q$ (red asterisks) is small as illustrated on two networks in Fig.~\ref{fig:stopcriterion}. Note that, this stop criterion is also applicable when network state oscillates among a group of states.

\begin{algorithm}
\caption{LabelRank}
\label{alg1}
\begin{algorithmic}[1]
\State add selfloop to adjacency matrix $A$
\State initialize label distribution $P$ using Eq. \ref{eq:initP}
\Repeat
	\State $P'=A \times P$
	\State $P'= \Gamma _{in} P'$
	\State $P'= \Phi_{r} P'$
	\State $P= \Theta _{q}(P',P)$
\Until{stop criterion satisfied}
\State output communities
\end{algorithmic}
\end{algorithm}	

\begin{figure*}[t]	
  \centering
	\includegraphics[width=12cm, height=6.5cm]{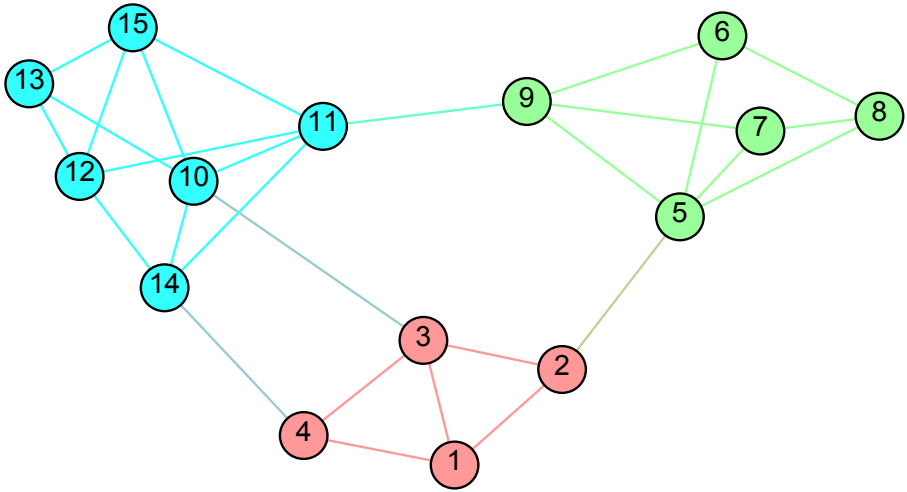}
	\caption{The example network G(0) with $n=15$. Colors represent communities discovered by LabelRank (see table \ref{table:tabelG0}) with cutoff $r=0.1$, inflation $in=4$, and conditional update $q=0.7$. The algorithm stopped at the $7^{th}$ iteration. The average number of labels dropped from 2.933 to 1.2 during the simulation.}
	\label{fig:G0}
\end{figure*}

\begin{table*}[t]
\centering
\caption{A sparse representation of the resultant matrix P on the example graph G(0) that defines probability of each label for each node. Note that for this matrix with $N=13$ nodes, there are at most two labels with non-zero probability for each node.}
\label{table:tabelG0}
\scalebox{1.1}{
\addtolength{\tabcolsep}{5pt}
\begin{tabular}{|c|c|c|c|c|} \hline
Node Identifier & $Label_1$ & $Probability_1$        & $Label_2$ & $Probability_2$\\ \hline
1  & \textbf{3}  & \textbf{0.721} & 1  & 0.279\\ \hline
2  & \textbf{3}  & \textbf{1}     &-   &-\\ \hline
3  & \textbf{3}  & \textbf{1}     &-   &-\\ \hline
4  & \textbf{3}  & \textbf{1}     &-   &-\\ \hline
5  & \textbf{5}  & \textbf{1}     &-   &-\\ \hline 
6  & \textbf{5}  & \textbf{1}     &-   &-\\ \hline 
7  & \textbf{5}  & \textbf{1}     &-   &-\\ \hline 
8  & \textbf{5}  & \textbf{1}     &-   &-\\ \hline 
9  & \textbf{5}  & \textbf{1}     &-   &-\\ \hline 
10 & \textbf{11} & \textbf{1}     &-   &-\\ \hline 
11 & \textbf{11} & \textbf{1}     &-   &-\\ \hline 
12 & \textbf{11} & \textbf{1}     &-   &-\\ \hline 
13 & \textbf{11} &\textbf{0.797}  & 10 & 0.203\\ \hline
14 & \textbf{11} & \textbf{1}     &-   &-\\ \hline 
15 & \textbf{11} &\textbf{0.874}  &10  &0.126\\ \hline
\end{tabular}
}
\end{table*}

These four operators together with a post-processing that groups nodes whose highest probability labels are the same into a community form a complete algorithm (see Alg.~\ref{alg1}). An example network as output by LabelRank is shown in Fig.~\ref{fig:G0}. There are only 1.2 labels on average and at most two in each node, resulting in a sparse label distribution (Table \ref{table:tabelG0} of which second row shows for each node the label with the highest probability identifying this node community). Three communities are identified, each sharing a common label: red community label 3, green community label 5 and blue community label 11. The resultant $P$ also distinguishes two types of nodes, the \textit{border} ones with high probability labels (e.g., 3, 5 and 11), and the \textit{core} nodes with positive but not largest label probabilities (e.g., 1, 13 and 15). The latter are well connected to their communities.  

In the analysis, we set the length of $P_i$ at $n$, creating a $n \times n $ $P$ matrix. In the implementation, this is not needed. Thanks to both cutoff and inflation operators, the number of labels in each node monotonically decreases and drops to a small constant in a few steps. The $P$ matrix is replaced by $n$ variable-length vectors (usually short) carried by each node (as illustrated in Table \ref{table:tabelG0}). Another advantage is that  the algorithm performance is not sensitive to the cutoff threshold $r$, so we set it to $0.1$, and do not consider when tuning parameters for optimal performance. 

It turns out that the preprocessing that adds a selfloop to each node (i.e., $A_{ii}=1$) helps to improve the detection quality. The selfloop effect resembles the lazy walk in a graph that avoids the periodicity problem, but here, it smooths the propagation (update of $P_i$) by taking into account node's own label distribution. Thus during initialization, each node considers itself a neighbor while using Eq. \ref{eq:bc}.

\begin{figure*}[t]	
  \centering
	\includegraphics[width=14cm, height=7.5cm]{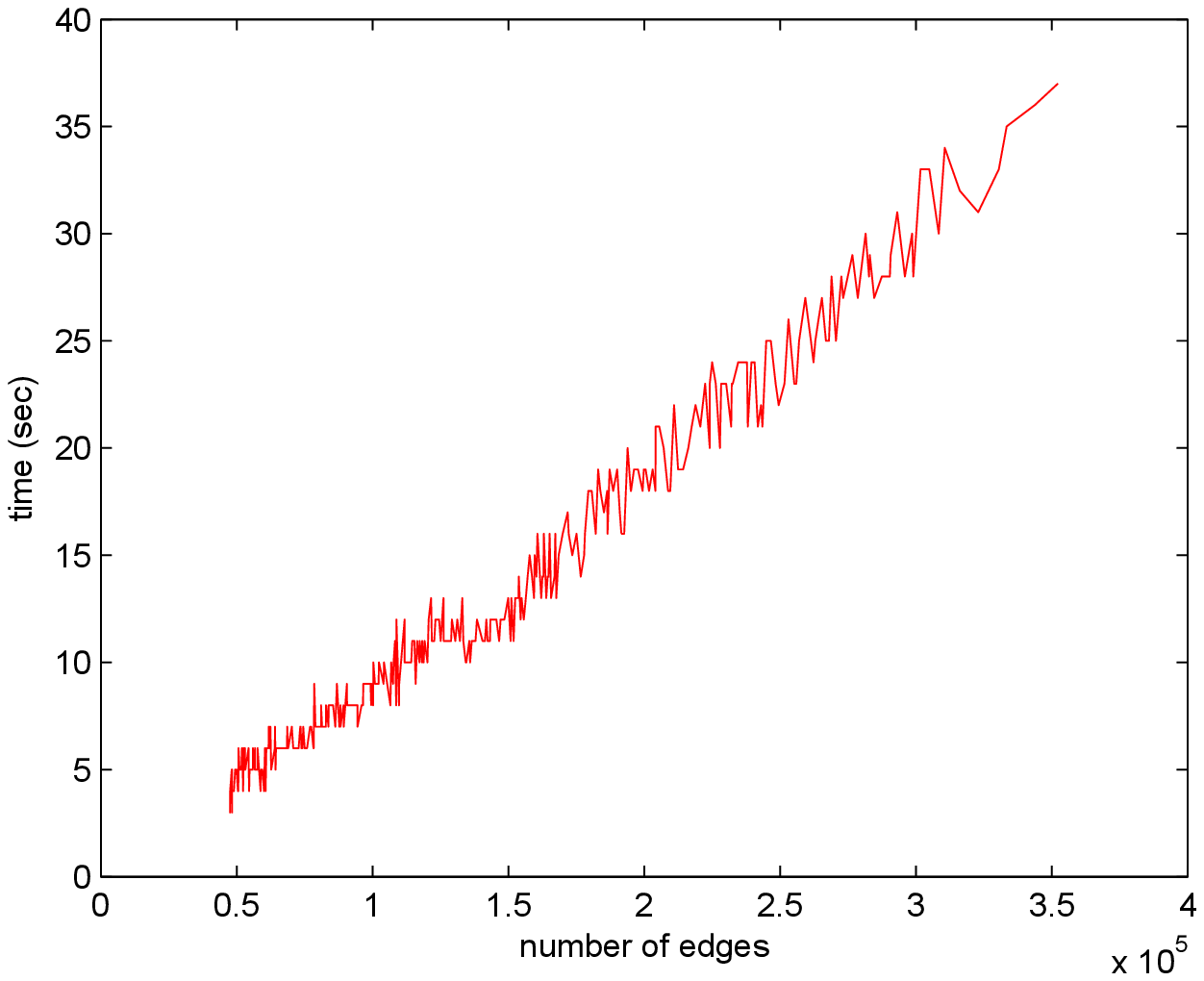}
	\caption{The execution times on a set of arXiv high energy physics theory citation graphs \cite{Leskovecdyn:2005} with  $n$ ranging from 12,917 to 27,769 and $m$ from 47,454 to 352,285. Tested on a desktop with Intel@2.80GHz.}
	\label{fig:runningTime-LabelRankOnly}
\end{figure*}

Both LabelRank and MCL use matrix multiplication, $A \times P$ for LabelRank and $M \times M$ for MCL (M is the $n \times n$ stochastic matrix). For updating an element, both $P_{ij} \leftarrow A_{i.} \times P_{.j}$ and $M_{ij} \leftarrow M_{i.} \times M_{.j}$ seem to require $O(n)$ operations, where $X_{i.}$ denotes the $i^{th}$ row and $X_{.j}$ denotes the $j^{th}$ column of matrix $X$. However, since $A$ represents the static network structure, no operations are needed for zero entries in $A$ for LabelRank. Thus, the number of effective operations for each node is defined by $k_i$ neighbors, reducing the time for computing the $P_{ij}$ to $O(k_i)$. With $x$ labels (typically less than 3) in each node on average, updating one row $P_i$ requires $O(xk_i)$ operations. As a result, the time for updating the entire $P$ in LabelRank is $O(x\overline{k}n)=O(xm)=O(m)$, where $\overline{k}$ is the average degree and $m$ is the total number of edges. In contrast, during the expansion (before convergence), $M_{ij}$ of $M$ that rises to power larger than 1 is changed according to the definition of transition matrix of a random walk. After that, values in $M_{ij}$ no longer reflects the network connections in one hop. Therefore, the computation of $M_{ij}$ may require nonlocal information and the time is $O(n)$, which leads to $O(nm)$ for the entire $M \times M$ operator in worst case. In conclusion, the propagation scheme in LabelRank is highly parallel and allows the computation to distribute to each individual node. 

The running time of LabelRank is $O(m)$, linear with the number of edges $m$ because adding selfloop takes $O(n)$, the initialization of $P$ takes $O(m)$, each of the four operators takes $O(m)$ on average and the number of iterations is usually $O(1)$. Note that, although sorting the label distribution is required in conditional update, it takes effectively linear time because the size of label vectors is usually no more than 3. The execution times on a set of citation networks are shown in Fig.~\ref{fig:runningTime-LabelRankOnly}. The test ran on a single computer, but we expect further improvement on a parallel platform. 

\begin{figure*}[t]
\centering	 
				\includegraphics[width=16cm, height=7.5cm]{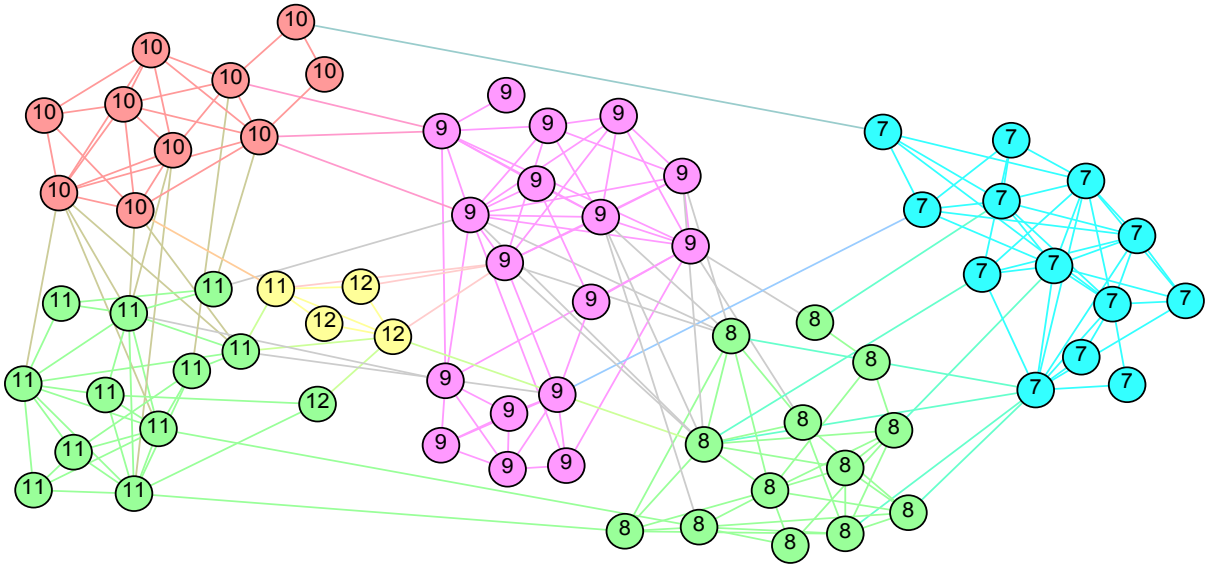}
				\caption{Communities detected on a HighSchool friendship network ($n=69$, $\overline{k}=6.4$). Labels are the known grades ranging from 7 to 12. Colors represent communities discovered by LabelRank.}
				\label{fig:comm1-LabelRank}
\end{figure*}	
\begin{table*}[t]
\centering
\caption{The modularity $Q$'s of different community detection algorithms.}
\label{table:compareQ}
\scalebox{1.1}{
\addtolength{\tabcolsep}{5pt}
\begin{tabular}{|c|c|c|c|c|c|} \hline
\textbf{Network} & \textbf{n} & \textbf{LPA} & \textbf{LabelRank} & \textbf{MCL} & \textbf{Infomap}	\\ \hline
Football  \cite{GirvanNewman:2002}& 115	&	0.60 &0.60 & 0.60	&	0.60\\ \hline
HighSchool & 1,127 & 0.66 & 0.66	&	0.60& 0.58\\ \hline
Eva & 4,475	& - & 0.89	& 0.89	& 0.89	\\ \hline
PGP \cite{dataPGP} & 10,680 & 0.63	& 0.81 & 0.80	&	0.81\\ \hline
Enron Email &	33,696 &	0.31 & 0.58	&	0.48 &	0.53\\ \hline
Epinions &	75,877 &	- & 0.34	& 0.27	&	0.38\\ \hline
Amazon \cite{dataAmazon}& 262,111	&	0.73 & 0.76	&	0.76 &	0.77\\ \hline
\end{tabular}
}
\end{table*} 	 

\section{Evaluation on Real-world Networks}
\label{sec:impdis}
We first verified the quality of communities reported by our algorithm on networks for which we know the true grouping. 
For the classical Zachary's karate club network \cite{Zac77} with $n=34$, 
LabelRank discovered exactly the two existing communities centered on the teacher and manager (with $Q=0.37$).

We also used a set of high school friendship networks \cite{JieruiXie-SLPA:2012} created by a project funded by the National Institute of Child Health and Human Development. The results on this large data set are similar and show
a good agreement between the found and known partitions. An instance is shown in Fig.~\ref{fig:comm1-LabelRank}.


We also tested LabelRank on a wider range of large social networks availbale at \url{snap.stanford.edu/data/} and compared its performance with other known algorithms including LPA with synchronous update \cite{Raghavan:2007}, MCL that uses a similar inflation operator \cite{MCL:2000} and one of the state-of-the-art algorithms, Infomap \cite{Santo:2010}. Since the output of LPA is nondeterministic, we repeated the algorithm 10 times and reported the best performance. For MCL, the best performance from inflation in the range of [1.5, 5] is shown. For LabelRank, $q$ is 0.5 or 0.6, $in$ is the best from the set $\{1,1.5,2\}$. Due to the lack of knowledge of true partitioning in most networks, we used modularity as the quality measure \cite{newman-2004-69}. The detection results are shown in Table. \ref{table:compareQ}.

As shown, LPA works well on only two networks with relatively dense average connections ($\overline{k}\approx10$): football and HighSchool networks. In general, it performs worse than the other three algorithms. However, with the stabilization strategies introduced in this paper, LabelRank, a generalized and stable version of LPA, boosts the performance significantly, e.g., with an increase of 28.57\% on PGP and  87.1\% on Enron Email. More importantly, LPA drawback is that it might easily lead to a trivial output (i.e., a single giant community). For instance, it completely fails on Eva and Epinions. The conditional update in LabelRank appears to provide a way to prevent such undesired output. As a result, LabelRank allows label propagation algorithms to work on a wider range of network structures, including both Eva and Epinions.

LabelRank outperforms MCL significantly on HighSchool, Epinions and Enron Email by 10\%, 20.83\% and 25.93\% respectively. This provides some evidence that 
there is an advantage of separating network structure captured in adjacency matrix $A$ from the label probability matrix $P$, as done in our LabelRank algorithm. LabelRank and Infomap have close performance. LabelRank outperforms Infomap on HighSchool and Epinions by 10.34\% and 9.43\% respectively, while Infomap outperforms LabelRank on Epinions by 11.76\%. 

\section{Conclusions}
In this paper, we introduced operators to stabilize and boost the LPA, which avoid random output and improve the performance of community detection. We believe the stabilization is important and can provide insights to an entire family of label propagation algorithms, including SLPA and COPRA. 

Stabilizing label propagation is our first step towards distributed dynamic network analysis. We are working on extending LabelRank for community detection for evolving networks, where new data come in as a stream. With such possible extension, we will be able to design efficient algorithms (e.g., distributed social-based message routing algorithm) for highly distributed and self-organizing applications such as ad hoc mobile networks and P2P networks. We also plan to extend LabelRank to overlapping community detection \cite{JieruiXie-Survey:2013} in the near future. In the experiments, we explored and demonstrated the good detection quality on some real-world networks. We are parallelizing our algorithm for millions of nodes networks.


\section*{Acknowledgment}

This work was supported in part by the Army Research
Laboratory under Cooperative Agreement Number
W911NF-09-2-0053 and by the Office of Naval Research
Grant No. N00014-09-1-0607. The views and
conclusions contained in this document are those of the
authors and should not be interpreted as representing
the official policies either expressed or implied of the
Army Research Laboratory, the Office of Naval Research, or the U.S. Government.


\end{document}